# Critical data analysis of COVID-19 spreading in Indonesia to measure the readiness of new-normal policy

**Muhammad ARIFUL FURQON [1,\*], Nina FADILAH NAJWA [2], Endah SEPTA SINTIYA [3], Eristya MAYA SAFITRI [4], and Iqbal RAMADHANI MUKHLIS [5]**

[1] Department of Information System, Institut Teknologi Sepuluh Nopember, Surabaya, Indonesia
[2] Department of Information System, Politeknik Caltex Riau, Pekanbaru, Indonesia
[3] Department of Information Technology, Politeknik Negeri Malang, Malang, Indonesia
[4] Department of Information System, Universitas Pembangunan Nasional Veteran Jawa Timur, Surabaya, Indonesia
[5] Department of Informatics, STIE Perbanas, Surabaya, Indonesia
E-mail(s): ariful.furqon16@mhs.is.its.ac.id[1\*], nina@pcr.ac.id[2], e.septa@polinema.ac.id[3], maya.si@upnjatim.ac.id [4], iqbal.ramadhani@perbanas.ac.id [5]
Tel(s): +62 878 0001 0033 [1\*]

# Critical data analysis of COVID-19 spreading in Indonesia to measure the readiness of new-normal policy

**Abstract**
COVID-19 pandemic has become a global issue nowadays. Various efforts have been made to break the chain of the spread of the COVID-19. Indonesia's government issued a large-scale social restrictions policy to prevent the spread of the COVID-19. However, large-scale social restrictions policy impacted the economy of the Indonesian. After several considerations, the Indonesian government implemented a new-normal policy, which regulates the activities outside the home with strict health protocols. This study's objective is to measure Indonesia's readiness level after the large-scale social restrictions period towards the new-normal period. To specify the readiness level, the measurement parameters required in the form of statistical analysis and forecasting modeling. Based on the results of statistical analysis and forecasting, over the past month, new confirmed cases increased more than two times. Besides, the growth rate of new confirmed cases dramatically increased rapidly compared to the prediction results. Therefore, the government must review the new-normal policy again and emphasize economic factors and think about health factors.

## Introduction

Coronavirus is a family of viruses with a spiky crown of glycoproteins on its surface [1]. The novel coronavirus, also known as COVID-19 or 2019-nCoV, is a contagious respiratory virus reported in Wuhan, China. On February 12th, 2020, the world health organization (WHO) designated the name was the coronavirus for the disease caused by the novel coronavirus as SARS-CoV-2 [2]. WHO has declared that COVID-19 is an essential international issue [3]. The COVID-19 pandemic spread to various parts of the world. Based on data obtained from John Hopkins University [4] as of June 5, 2020, the confirmed cases of COVID-19 in the World were 7,487,676, and the number of death cases was 420,236. Breaking the chain of the spread of this pandemic must be done seriously by each country.

Indonesia is also trying to break the COVID-19 spreading chain. Some countries use different methods to inhibit the spreading of COVID-19. Indonesia's government policy to inhibit the spreading of COVID-19, known as large-scale social restrictions, is a policy to reduce the spreading of the COVID-19 in Indonesia [5]. The large-scale social restriction has already become government policy based on some considerations [6]. Under government regulation number 21 of 2020, concerning large-scale social restrictions in the framework of accelerating the countermeasure of COVID-19 spreading [7]. The large-scale social restrictions aims are to anticipating and reducing the number of COVID-19 cases in Indonesia that carried out in all regions in Indonesia. The large-scale social restriction is to limit people's activities outside the house. These social restrictions are useful for suppressing the increase in new confirmed cases in Indonesia.

After about two months undergoing large-scale social restrictions, the Indonesian government issued a new-normal policy. The new-normal policy adopted by the Indonesian government is the result of the process of people adaptation during the COVID-19 pandemic [7]. The new-normal policy expected to be able to restore the economy, which has profoundly declined when large-scale social restrictions implemented. Although the new-normal policy carried out, vigilance against COVID-19 still needs to be done, such as the application of health protocols in the companies. The intended health protocol forced to maintain personal hygiene and the work environment, i.e., using masks, washing hands, and implementing physical distancing in every activity. The readiness of the

Indonesian government to implement new-normal life needs to be measured. The redesign of this new-normal life is needed so that this policy lasts [8].

The measurement of Indonesia's readiness to implement a new-normal policy after the large-scale social restrictions period is needed. Therefore, further studies to measure the level of readiness is needed. A statistical analysis study is needed to interpret the spreading of COVID-19 data [9]. Another study is to predict the spreading of COVID-19 in the future by using the forecasting model. Thus, this study's objective is to measure Indonesia's readiness level after the large-scale social restrictions period towards the new-normal period. To specify the level of readiness, the measurement parameters are needed in statistical analysis and forecasting modeling. Therefore, this study conducted statistical analysis techniques and forecasting modeling to observe the trends of the spreading of COVID-19 after the large-scale social restrictions period. This study's expected results can provide insight, especially to the Indonesian government, to take the future policy to countermeasure the spreading of COVID-19 in Indonesia.

**Related Works**

The COVID-19 issue hotly discussed, in both the medical and non-medical fields. COVID-19 countermeasures require a comprehensive study from various perspectives of knowledge and not only from medical knowledge. Various studies are currently being carried out by researchers from various scientific fields around the world.

A forecasting study of the COVID-19 outbreak conducted by [10] predicts the average total cumulative case count in Hubei by February 24, 2020, using susceptible-infectious-recovered-dead (SIDR) model. A study from [11] used phenomenological models that have been validated during previous outbreaks to generate and assess short-term forecasts of the cumulative number of confirmed reported cases in Hubei province. The research about modeling simulation in Indonesia conducted by [12] using Richard's Curve that represents a modified logistic equation. Research conducted by [13] regarding the forecasting of COVID-19 transmission and healthcare capacity in Bali, Indonesia. The forecast provided an understanding of the effectiveness of current intervention strategies and responses to case handlers. The COVID-19 data is a time-series data set that extracted to observe the spreading [14]. Therefore, a procedure for forecasting time series data is needed. A study conducted by [15] used the Facebook prophet model to forecast COVID-19. Research in 2020 by [16] using the Logistic and Facebook Prophet models to forecast COVID-19 data trends from global cases and several countries as samples. The Indonesian state is one of the forecasting samples, but it has not explicitly explained the phenomenon and the implementation of government policy in the COVID-19 case in Indonesia.

Indonesia's government policy regarding the countermeasure of COVID-19 was also examined by [6]. The study results showed that Indonesia has quite a significant concern for the COVID-19 phenomenon. Thus, government policy in implementing large-scale social restrictions needed as an effort to deal with the spreading of the COVID-19. On the contrary, large-scale social restriction policies in Indonesia prevented people from working outside the home. One of the impacts of large-scale social restrictions on Indonesian was the termination of employment. The study conducted by [17] shown that COVID-19 impacted the increasing poverty rate in Indonesia. To reduce the poverty rate in Indonesia because of COVID-19, the government issued a new-normal policy to do activities outside the home with strict health protocols. The new-normal policy with mitigation actions aimed to reduce or minimize the impact of COVID-19 in Indonesia, but the people can still work outside the home [7]. The steps taken by Indonesia's government in implementing new-normal were to maintain economic growth and financial stability [18]. Thus, some economic activity in Indonesia can still be carried out even with strict and massive restrictions.

**Methodology**

The study conducted comprehensively to obtain information that can provide insight. Therefore,

a methodology was needed to obtain and analyze COVID-19 data illustrated by a block diagram presented in Figure 1.

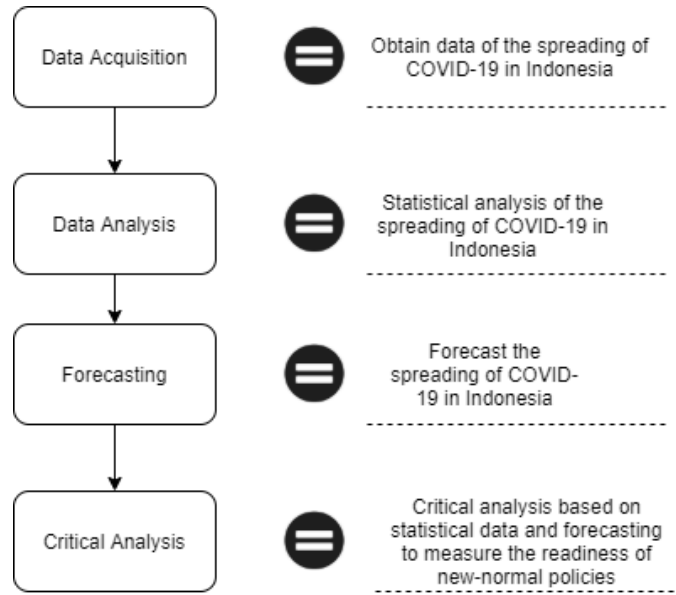

Figure 1. Methodology block diagram

The first step carried out in this study is data acquisition. Data acquisition conducted to obtain data on the spreading of COVID-19 in Indonesia. After all the data acquisition processes have been carried out and the COVID-19 data distribution CSV file obtained, the next process is data analysis. In addition to data analysis, forecasting using Facebook Prophet also carried out to project the spreading of COVID-19. Based on data analysis and forecasting, critical analysis carried out to measure government readiness in implementing new-normal policy.

*Data*

The study was based on the COVID-19 data provided by Johns Hopkins University [4]. The data obtained from Johns Hopkins University extracted to obtain the spreading of COVID-19 data in Indonesia. The data obtained includes (1) date, (2) number of confirmed cases; (3) number of deaths; and (4) number of recovery. The data used in this study limited from March 2, 2020, to June 5, 2020. The limitation of the data used is because the first confirmed case of COVID-19 that occurred in Indonesia on March 2, 2020 [19] and most regions in Indonesia ended large-scale social restrictions period on June 5, 2020.

*Methods*

Statistical analysis is an essential component of all medical informatics research to uncovering patterns and trends of some phenomena. The use of descriptive and inferential methods enables researchers to summarize findings and conduct hypothesis testing [20]. Thus, several statistical analyses carried out to gain insight and trends from the spreading of COVID-19 data in Indonesia.

The first statistical analysis that discussed is the calculation of the number of confirmed cases *(N confirmed)*, recovered cases *(N recovered)*, and death cases *(N death)* from COVID-19 patients in Indonesia. The subsequent analysis is the calculation of the number of active cases and closed cases. The number of active cases is the number of confirmed cases reduced by the number of other cases and can be formulated in equation 1.

$$Active\ cases\ =\ N\ confirmed\ -\ N\ recovered\ -\ N\ death \qquad (1)$$

An increase in the number of active cases is an indication of recovered cases or death cases number is decreasing compared to the number of confirmed cases increasing drastically. Conversely, if the number of an active case decreases, it indicates that recovered, and death cases are increasing, and fewer new confirmed cases. On the other hand, the number of closed cases represents the total number of recovered and death cases. The number of closed cases can be formulated in equation 2.

$$Closed\ cases\ =\ N\ recovered\ +\ N\ death \qquad (2)$$

The increase in the number of closed cases implies that more patients are getting recovered or died. The next statistical analysis parameters are the mortality rate and recovery rate that formulated with equations 3 and 4.

$$Mortality\ rate\ =\ \frac{N\ death}{N\ confirmed} \times 100 \qquad (3)$$

$$Recovery\ rate\ =\ \frac{N\ recovered}{N\ confirmed} \times 100 \qquad (4)$$

The mortality rate shows a measure of the frequency of occurrence of death in a defined population during a specified interval [21]. The lower in the mortality rate indicates the better COVID-19 countermeasures. The opposite of the mortality rate is the recovery rate that shows a measure of the frequency of patients' recovery in a defined population during a specified interval [21].

Because COVID-19 data is a time-series data [14], this study used the Facebook prophet model that is a procedure for forecasting time series data based on an additive model where non-linear trends are fit with daily seasonality and holiday effects [15]. The Facebook prophet model works best with time-series data, with strong seasonal effects and several seasons of historical data. It is also robust to missing data and shifts in the trend. For the average method, the forecasts of all future values are equal to the average or mean historical data. Let y1 denote the historical data, $\ldots, y_T$, and the prophet forecast can be formulated with equation 5.

$$\hat{y}_T\ +\ h|T\ =\ \bar{y}\ =\ \frac{(y_1\ +\ y_2 + \ldots + y_T)}{T} \qquad (5)$$

$\hat{y}_T\ +\ h|T$ is a short-hand for the estimated value of $y_T\ +\ h$ based on the historical data $y_1, \ldots, y_T$. A prediction interval gives an interval within expected $y_t$ to lie with a specified probability. Assuming that the forecast errors normally distributed, a 95% prediction interval for the $h$-step forecast can formulate with equation 6.

$$\hat{y}_T\ +\ h|T\ \pm 1.96\ \hat{\sigma}_h \qquad (6)$$

where $\hat{\sigma}_h$ is the estimate of the standard deviation of the $h$-step forecast distribution.

*Statistical Analysis and Forecasting Modelling*

The statistical analysis was made by using python 3.7 programming language. As explained in the method sub-section, this study utilized forecasting models adapted to overcome the complexity of COVID-19 time series data. The forecasting modeling also carried out using libraries available in python. The statistical analysis and forecasting results presented in the results section.

**Results and Discussion**

As discussed in the previous section, this study using statistical analysis and forecasting modeling to gain insight into the future trends of the spreading of COVID-19. The statistical analysis and forecasting modeling of the spreading of COVID-19 in Indonesia presented in the following sub-sections.

*Statistical Analysis*

The statistical analysis aims to identify the trends in the spreading of COVID-19. The statistical analysis carried out on the spreading of COVID-19 data in Indonesia from March 2, 2020, until June 5, 2020. Based on the analysis that has been done, the data obtained as shown in Table 1

**Table 1.** Spreading of COVID-19 data in Indonesia statistical analysis

| Parameter(s) | Total |
|---|---|
| Total number of confirmed cases | 29,521 |
| Total number of recovered cases | 9,443 |
| Total number of deaths cases | 1,770 |
| Total number of active cases | 18,308 |
| Total number of closed cases | 11,213 |
| The approximate number of confirmed cases per day | 308 |
| The approximate number of recovered cases per day | 98 |
| The approximate number of death cases per day | 18 |
| The approximate number of confirmed cases per hour | 13 |
| The approximate number of recovered cases per hour | 4 |
| The approximate number of death cases per hour | 1 |
| Average mortality rate | 6.6994 |
| Median mortality rate | 7.3105 |
| Average recovery rate | 12.8777 |
| Median recovery rate | 10.3408 |

Based on Table 1, the total confirmed, recovered, and death cases from June 5, 2020, were 29,521, 9,443, and 1,770, respectively. The number of active and closed cases was 18,308 (62%) and 11,213 (38%). Based on closed cases, it can imply that the overall recovered and mortality rates were 84% and 16%, respectively. The approximate number of confirmed, recovered, and death cases per day imply that there was an average of 308 new confirmed, 98 recovered, and 18 death cases every day. The approximate number of confirmed, recovered, and death cases per hour imply that there was an average of 13 new confirmed, four recovered, and one death cases every hour. The average and median mortality rates were 6.6994 and 7.3105, respectively. Otherwise, the average and median of recovery rate were 6.6994 and 7.3105, respectively.

Analysis of COVID-19 spreading data distribution performed to show all of the cases (confirmed, recovered, and death) day-to-day. The distribution plot of COVID-19 in Indonesia presented in Figure 2. The analysis of data distribution can be used to analyze the growth of all COVID-19 cases in Indonesia. The growth of different types of cases in Indonesia presented in Figure 3. The number of weekly increases cases COVID-19 is required to analyze the growth trend in the number of new cases of COVID-19. The weekly increase in the number of confirmed cases and death cases presented in Figure 4. Two parameters that are indicators of the success or failure of COVID-19 countermeasure are recovery and death rate. The higher recovery rate and lower mortality rate are indicators of success in overcoming COVID-19, and so otherwise. The recovery rate and mortality rate of COVID-19's patients in Indonesia shown in Figure 5.

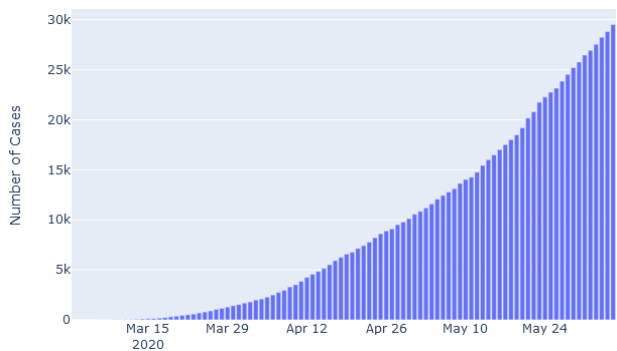
(a)

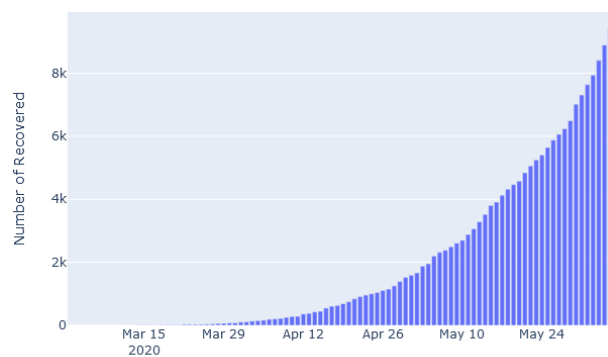
(b)

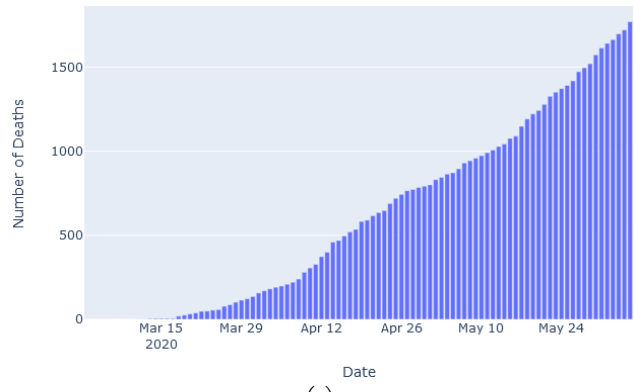

(c)

**Figure 2.** The distribution of **(a)** number of cases, **(b)** number of recovered, and **(c)** number of deaths in Indonesia (up to June 5, 2020)

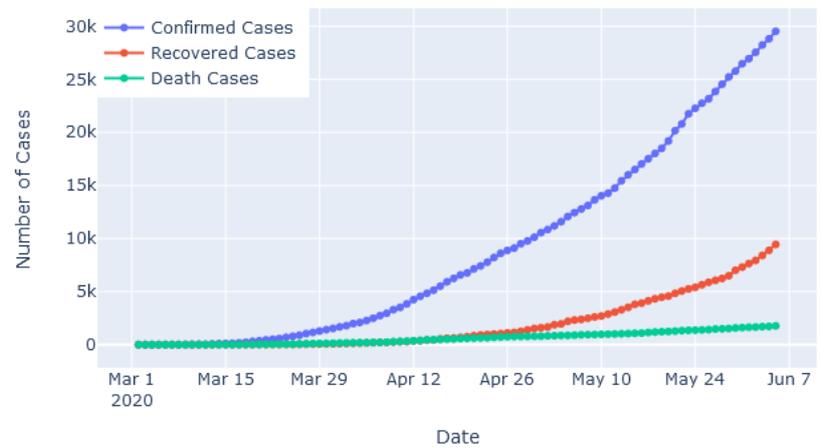

**Figure 3.** The growth of different cases (confirmed, recovered, and death) in Indonesia (up to June 5, 2020)

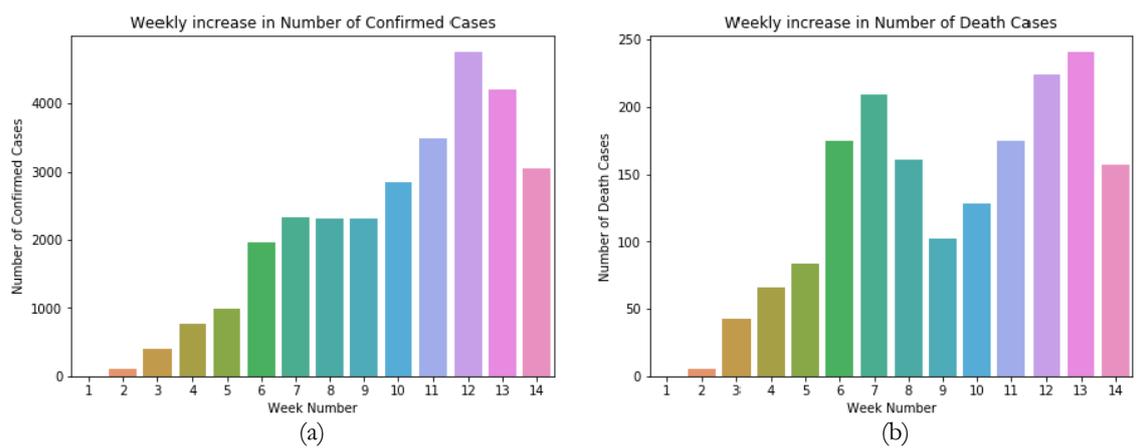

(a) (b)

**Figure 4.** The weekly increase in the number of **(a)** confirmed cases and **(b)** death cases in Indonesia (up to June 5, 2020)

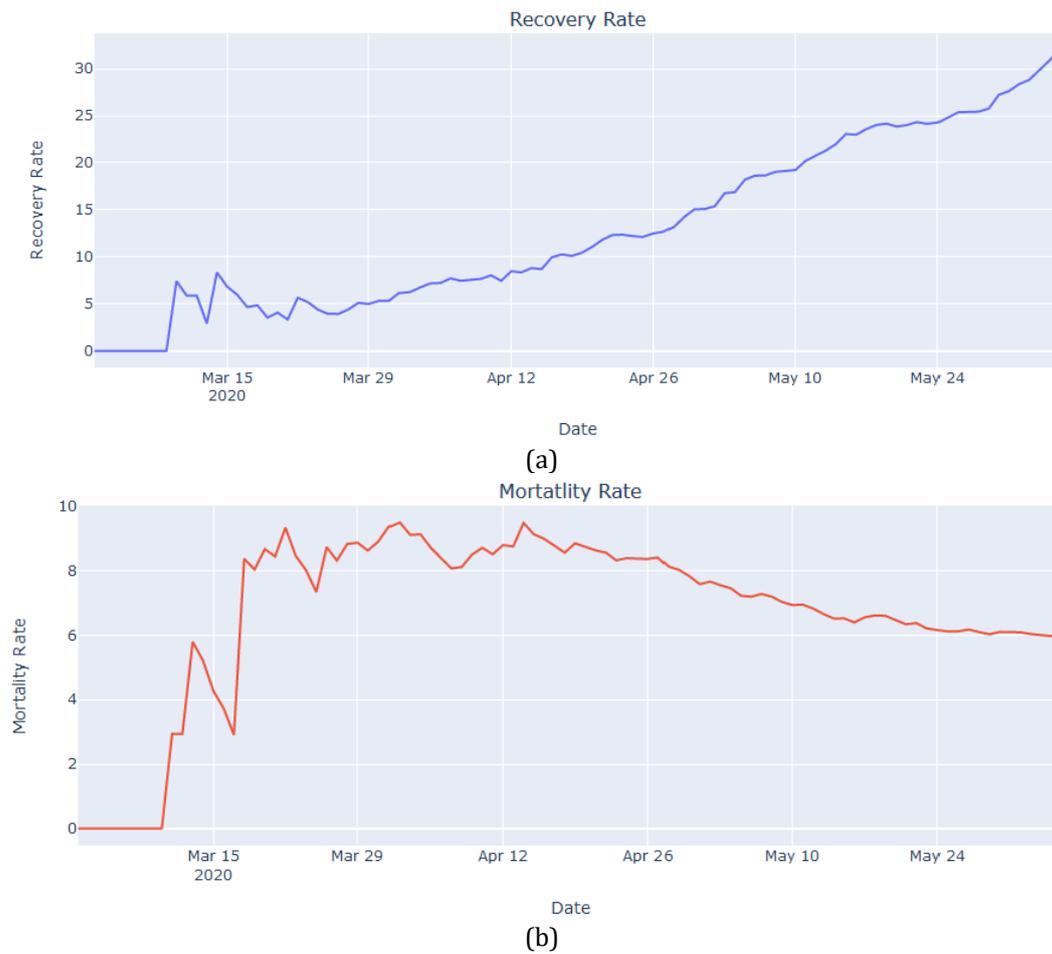

**Figure 5.** The **(a)** recovery rate and **(b)** mortality rate of COVID-19 patients in Indonesia (up to June 5, 2020)

Figures 2 and 3 describe the distribution and growth of the number of confirmed, recovery, and death cases. The number of confirmed cases, growth, recovery, and death will increase from time to time. Based on Figure 4a, the number of confirmed cases increases from week to week. The highest increase in the number of new confirmed cases occurred in the 12th week. The number of new confirmed cases has decreased from the 12th week to the 14th week. Thus, it can be concluded that there is an impact of large-scale social restrictions on the growth of new confirmed cases. Based on Figure 4b, the death number each week has fluctuated during the time, where the peak occurred in the 13th week. However, in the 14th week, the number of death cases has decreased, which is a good sign. Based on Figure 5, the mortality rate before the implementation of the new-normal policy showed a considerable drop for a pretty long time, which is a positive sign. The recovery rate has started to pick up again, which is a good sign, another supportive reason why the number of closed cases is increasing.

*Forecasting*

This study performed prediction based on COVID-19 confirmed cases in Indonesia during the first confirmed case (March 2, 2020) until the implementation of the new-normal policy (June 5, 2020). Facebook prophet model handles missing data and outliers of the trend [22]. Facebook prophet model also used as a procedure for forecasting time series data based on an additive model where non-linear trends are fit with yearly, weekly, and daily seasonality [15]. Thus, using the Facebook prophet modeled COVID-19 confirmed cases in Indonesia. Figure 5a shows an increasing trend in confirmed cases of COVID-19 after the implementation of the new-normal policy. Whereas Figure 5b shows the day of week seasonality trend on the increasing confirmed cases.

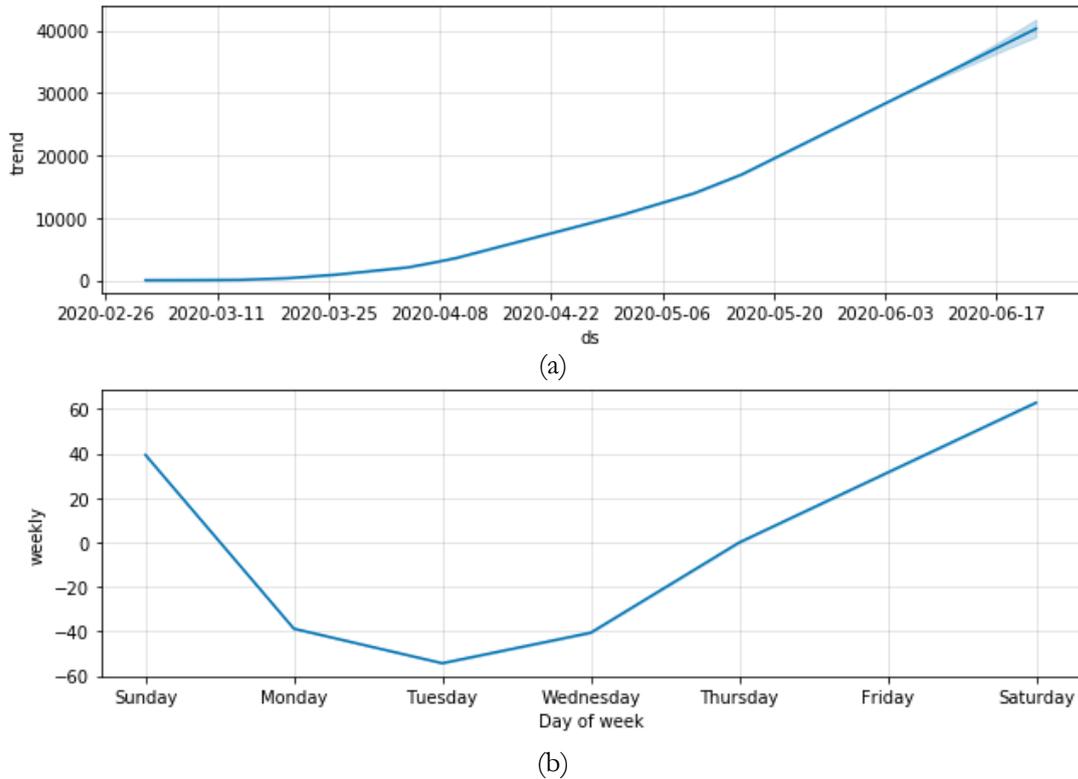

(a)

(b)

**Figure 6.** The trend of (a) the increasing confirmed cases and (b) day of week seasonality using the Facebook Prophet model

Based on the trend of increasing confirmed cases, the number of confirmed cases can be forecasted in the future. In this study, the forecasting was performed for one month from June 6, 2020, until July 5, 2020. Besides, a comparison between the prediction results and the actual confirmed cases performed, as described in Table 2. Moreover, a plot presented the predicted and actual confirmed cases after the implementation of the new-normal policy, as presented in Figure 7.

**Table 2.** Comparison of the predictions and actual confirmed cases from June 6 until July 5

| Dates | Prophet Model Prediction | Actual Confirmed Cases |
|---|---|---|
| June 6, 2020 | 30,561 | 31,186 |
| June 7, 2020 | 31,265 | 32,033 |
| June 8, 2020 | 31,987 | 33,076 |
| June 9, 2020 | 32,709 | 34,316 |
| June 10, 2020 | 33,446 | 35,295 |
| June 11, 2020 | 34,183 | 36,406 |
| June 12, 2020 | 34,932 | 37,420 |
| June 13, 2020 | 35,686 | 38,227 |
| June 14, 2020 | 36,450 | 39,294 |
| June 15, 2020 | 37,220 | 40,400 |
| June 16, 2020 | 38,000 | 41,431 |
| June 17, 2020 | 38,785 | 42,762 |
| June 18, 2020 | 39,579 | 43,803 |
| June 19, 2020 | 40,380 | 45,029 |
| June 20, 2020 | 41,189 | 45,891 |
| June 21, 2020 | 42,006 | 46,845 |
| June 22, 2020 | 42,830 | 47,896 |
| June 23, 2020 | 43,662 | 49,009 |
| June 24, 2020 | 44,502 | 50,187 |
| June 25, 2020 | 45,349 | 51,427 |
| June 26, 2020 | 46,204 | 52,812 |
| June 27, 2020 | 47,066 | 54,010 |
| June 28, 2020 | 47,936 | 55,092 |
| June 29, 2020 | 48,814 | 56,385 |
| June 30, 2020 | 49,699 | 57,770 |
| July 1, 2020 | 50,592 | 59,394 |
| July 2, 2020 | 51,493 | 60,695 |
| July 3, 2020 | 52,401 | 62,142 |
| July 4, 2020 | 53,317 | 63,749 |
| July 5, 2020 | 54,241 | 64,958 |

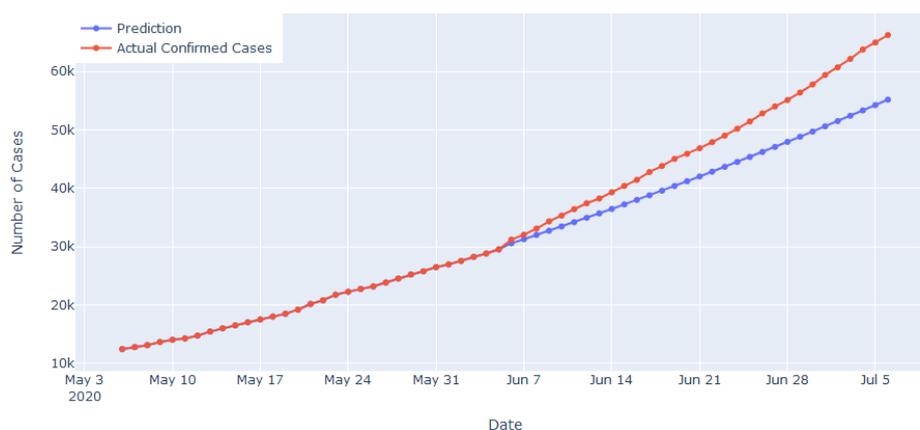

**Figure 7.** Comparison of predicted and actual confirmed cases related to the implementation of the new-normal policy

Based on the statistical analysis results, the total confirmed cases up to June 5, 2020, were 29,521. Compared to actual confirmed cases in Table 2, the confirmed cases up to July 5, 2020, increased into 64,958. It implies that over the past month, new confirmed cases increased by more than two times. Based on Figure 7, the growth rate of new confirmed cases dramatically increased rapidly compared to the prediction results. It indicates the bad sign related to the implementation of the new-normal policy.

*Critical Analysis*

Two months after the implementation of the large-scale social restrictions, the Government of Indonesia is now preparing to roll back sluggish Indonesia's economy due to the implementation of the large-scale social restrictions. Indonesia's government aims to adapt new-normal policy is to move up the people's economy alongside the COVID-19 pandemic. Many researchers consider that new-normal policies have positive and negative impacts. The enactment of this policy can make people vulnerable to contracting COVID-19. However, if this policy did not implement, it can force people to stay at home, which will have a massive economic impact. After going through several considerations, the government implemented a new-normal policy.

Unfortunately, right after the adaptation of new-normal policy in several regions, the daily confirmed cases were increased dramatically. It reinforced by the increase in the number of new confirmed cases by more than two times. The growth rate of new confirmed cases also dramatically increased rapidly compared to the prediction results over the past month. It indicates the bad sign related to the implementation of the new-normal policy. Therefore, the government must review the new-normal policy again, in addition to emphasizing economic factors must also think about health factors.

To suppress the growth rate of new confirmed cases, the government must be assertive in implementing new-normal policies. The government also must carry out massive socialization towards the adaptation of new-normal policy. Besides, people also should be aware of complying with health protocols that have been made by the government. The majority of new confirmed cases were occurring due to people's negligence in complying with the policy. Hence, people must also be actively involved in breaking the chain of COVID-19 spreading in Indonesia by complying with the health protocol. Lastly, if the new-normal wants to be successful in carrying out new-normalcy, all stakeholders and the public must be disciplined in carrying out the health protocols and rules that have been made.

**Conclusions and Future Works**

The COVID-19 pandemic spread to various parts of the world, including Indonesia. The various strategies had been implemented to prevent the spreading of the coronavirus. Indonesia's government policy/strategy to inhibit the spreading of COVID-19, known as large-scale social restrictions, is a policy to reduce the spreading of the COVID-19 in Indonesia. After the large-scale social restrictions, the government policy was implemented in the new-normal period. As long as the new normal period, the government gave the instructions to citizens for implementing the health protocol in every environment, including in-office and market place. The parameters for the success or failure of COVID-19 countermeasures are recovery and death rate. The recovery rate and mortality rate of COVID-19's patients in Indonesia, before the implementation of the new-normal policy, was showing a considerable dropping for a pretty long time, which is a positive sign. This study aimed to assess Indonesia's readiness level towards the new-normal period. The readiness level that can be assessed by using the Covid-19 data extracted.

The Facebook prophet model works best with time-series data. It is also compelling for data loss and trend shifts, so using this method is the right step. After the Facebook prophet forecasting method was conducted for one month from 6 June 2020, until 5 July 2020, the growth rate of new confirmed cases dramatically increased rapidly compared to the predicted results. The growth rate shows a lousy sign related to the implementation of new-normal policies. However, the government continues to implement new-normal policies after going through various considerations, one of which is the economic impact. Based on the results of statistical analysis and forecasting, over the past month, new confirmed cases increased more than two times. Most of the confirmed new cases occur due to people's negligence in complying with the policy.

This study has several limitations which open up directions for future research. Firstly, The data used in this study is static data for each province in Indonesia. Future research can use the dynamic data confirmed cases of COVID-19, so the value of the resulting forecasting more up to date. Secondly, this study mainly concentrates on measure Indonesia's level of readiness after the large-scale social restrictions period towards the new normal period. Future studies can conduct forecasting analysis related to the economy of each province affected by the COVID-19 pandemic. The results of the forecast can be used to explore aspects that need to be improved or avoided to mitigate the economic downturn in the next few years due to the case of COVID-19. Thirdly, Future studies can use forecasting methods in addition to using the Facebook prophet and comparing the accuracy of the values generated.

**List of abbreviations (if any)**

| | |
|---|---|
| COVID-19 | Coronavirus disease 2019 |
| CSV | Comma-Separated Values |
| RMSE | Root Mean Square Error |

**Conflict of Interest**

*The author(s) declare that they have no conflict of interest'.*

**Authors' Contributions**

Muhammad Ariful Furqon defined the aim of research and the design of the experiment. He also conducted data acquisition to obtain COVID-19 data in Indonesia. Nina Fadilah Najwa contributed to conducting studies and analysis related to the spreading of COVID-19 and its countermeasures. Endah Septa Sintiya assisted in analyzing COVID-19 forecasting modeling. Muhammad Ariful Furqon, with the help of Iqbal Ramadhani Mukhlis and Eristya Maya Safitri, were conducting statistical analysis and forecasting modeling. All authors participate in coordinate and assist in draft the manuscript. All authors have read and approved the final manuscript.